\documentstyle[11pt,newpasp,twoside,graphicx]{article}
\def\edcomment#1{\iffalse\marginpar{\raggedright\sl#1\/}\else\relax\fi}
\reversemarginpar
\def\ls{{_<\atop^{\sim}}}

\def\cgs{ ${\rm erg~cm}^{-2}~{\rm s}^{-1}$ } 

\begin{document}
\title{The AGN content of hard X-ray surveys}
 \author{Andrea Comastri}
\affil{Osservatorio Astronomico di Bologna, via Ranzani 1, I--40127 Bologna,
  Italy}
\author{Cristian Vignali}
\affil{Dept. of Astronomy and Astrophysics, Penn State University, 525 Davey Lab -- University Park, PA 16802, USA}
\author{Marcella Brusa}
\affil{Dipartimento di Astronomia, Universita' di Bologna, via Ranzani 1, 
I--40127 Bologna, Italy}  
\author{on behalf of the HELLAS and HELLAS2XMM consortii\footnotemark[1]}
\footnotetext[1]{HELLAS:
F. Fiore, P. Giommi, G. Matt, F. La Franca, G.C. Perola, 
S. Molendi, R. Maiolino, A. Antonelli; 
HELLAS2XMM: F. Fiore, A. Baldi, S. Molendi, M. Mignoli, P. Ciliegi, 
F. La Franca, G. Matt, G.C. Perola, P. Severgnini, R. Maiolino}

\begin{abstract}

Multiwavelength observations of the hard X--ray selected sources 
discovered by {\it BeppoSAX}, {\it Chandra} and XMM--{\it Newton} surveys 
have significantly improved our knowledge of the 
AGN population. 
The increasing number of X--ray obscured AGN 
so far discovered confirms the prediction of those AGN synthesis 
models for the X--ray background based on the Unified scheme.
However, follow--up optical observations of hard X--ray selected sources
indicate that their optical properties are quite varied
and the simple relations between optical and X--ray absorption
are by no means without exception.
Moreover there is evidence of a substantial number 
of luminous X--ray sources hosted by apparently normal 
galaxies. 
In this paper the results obtained from multiwavelength 
observations of hard X--ray selected sources discovered by 
{\it BeppoSAX} and XMM--{\it Newton} are presented and briefly discussed.

\end{abstract}

\section{Introduction}

A large fraction of the energy density contained 
in the cosmic X--ray background spectrum (XRB) 
is accounted for by the summed contribution 
of Active Galactic Nuclei (AGN) if most of their high-energy radiation 
integrated over the cosmic time is obscured by gas and dust.
Several independent models based on AGN unification scheme 
(Setti \& Woltjer 1989, Madau et al. 1994, Comastri et al. 1995,
Gilli et al. 2001) 
and energetic arguments (Fabian \& Iwasawa 1999)
lead to the conclusion that a fraction as high as 80--90\%
of the luminosity produced by accretion-powered sources 
is obscured at almost all wavelengths
emerging only in the hard X--ray band above a few keV.
Hard X--ray surveys represent thus the most efficient method to search for 
and to trace the cosmological evolution of accretion-powered sources.
A still unknown fraction of obscured radiation is reprocessed and 
remitted in the far--infrared band.
Multiwavelength follow--up observations of 
hard X--ray selected sources would allow to probe 
where the bulk of obscured accretion power is remitted 
and estimate the AGN contribution to the extragalactic background light
in the infrared band. 

Thanks to their revolutionary capabilities (arcsec imaging 
and high-energy throughput)
{\it Chandra} and XMM--{\it Newton} have opened up a new era in the study 
of the hard X--ray sky. 
Deep {\it Chandra} surveys (Brandt et al. 2001, Rosati et al. 2001)
have reached extremely faint fluxes in the 0.5--2 keV and 2--7 keV bands
virtually resolving the entire XRB flux at these energies;
while relatively deep XMM--{\it Newton} 
exposures (Hasinger et al. 2001, Baldi et al. 2001) 
have extended by a factor 50 the sensitivity in the 2--10 and 5--10 
keV bands with respect to previous {\it ASCA} and {\it BeppoSAX} observations.
The X--ray source counts and their average spectral properties, which 
are now probed over several decades of fluxes and energy ranges, 
appear to be consistent with AGN synthesis model predictions.
Although the remarkable results obtained so far, deep {\it Chandra} 
and XMM--{\it Newton} surveys are limited by small area coverage 
(less than a quarter of square degree) and by 
the extremely faint magnitudes of the optical counterparts 
which make the identification of X--ray sources very difficult, 
if not impossible, even at 8m class telescopes 
(Giacconi et al. 2001, Tozzi et al. 2001, Norman et al. 2001).
In order to fully characterize the nature and 
evolution of the X--ray source population it is customary 
to complement deep surveys with shallower observations  
carried out on larger areas (see for example the Einstein Medium Sensitivity
Survey: EMSS; Gioia et al. 1990). 
This approach allows us to minimize the effects of field-to-field 
fluctuations (the cosmic variance) and makes much easier 
the optical identification follow-up observations given the 
average brighter magnitude of the counterparts.

In this review the results obtained by 
two Large area surveys carried out with {\it BeppoSAX} ({\tt HELLAS})
and XMM--{\it Newton} ({\tt HELLAS2XMM}) are summarized.
The main scientific drive of this project is to probe the
obscured accretion history of the X--ray Universe. The adopted 
strategy is a trade--off between the hardest X--ray band 
and the largest area which can be covered with {\it BeppoSAX} 
and XMM--{\it Newton}.

\section{The HELLAS view of the hard X--ray source population}

The High Energy Large Area Survey ({\tt HELLAS}; Fiore et al. 1999, 2001a) 
has provided, for the first time before the advent of the new X--ray 
observatories {\it Chandra} and XMM--{\it Newton}, a well-defined and 
large-area sample of hard (5--10 keV) X--ray selected sources 
obtained with an imaging instrument (the MECS instrument onboard 
{\it BeppoSAX}). 
At the flux limit of about 5 $\times$ 10$^{-14}$ ${\rm erg~cm}^{-2}~{\rm s}^{-1}$, 
the integrated flux of the HELLAS sources account for some 20--30 \% of the 
hard 5--10 keV XRB flux (Comastri et al. 2001).

The optical identification follow--up has been carried out on a 
subsample of 118 HELLAS sources covering $\sim$ 55 deg$^2$.
The details of sample selection and optical identification breakdown are 
reported in Fiore et al. (2001a, 2001b) and La Franca et al. (2002).
About 60\% of the objects have been optically identified either 
by cross--correlation with public catalogs (25 objects) 
or at 4m class telescopes (49 objects). 
For 13 sources there are no clear counterparts down to R$\simeq$21.
Even if the optical identification are dominated by type 1 
AGN, the relative fraction of type 2 objects 
(including Seyfert~1.8, 1.9 and emission--line galaxies)
vs. type~1 is about 0.40, 
considerably higher than the value of 0.25 found in the {\it ROSAT} 
Ultradeep Survey (Lehmann et al. 2001a) and that of 0.20 
in the {\it ASCA} Large Area Survey (LSS; Akiyama et al. 2000). 
If one assumes that the 13 ``empty'' fields 
host Type~2 AGNs, then the Type~2/Type~1 ratio becomes of about 0.7. 
This shows the efficiency of revealing obscured X--ray sources 
in the 5--10 keV band with {\it BeppoSAX}. 

Although for the majority of the HELLAS sources a ``standard'' 
spectral analysis is not possible due to the low photon statistics, 
the analysis of X--ray colors (hardness ratios) allows us 
to unambiguously reveal a substantial number of
objects with flat and/or absorbed X--ray spectra.
The most intriguing result is the presence of hard, presumably 
absorbed X--ray spectra in objects optically classified as type 1
AGN with a blue optical/UV continuum and broad lines. The 
number of X--ray absorbed type 1 quasars appears to increase 
with redshift and/or luminosity (Fig.~1).

%%%%%
\begin{figure}[!h]
%\plotone{fig1.eps}
\centerline{\includegraphics[angle=0,width=\textwidth,bbllx=20pt,bblly=15pt,bburx=550pt,bbury=400pt]{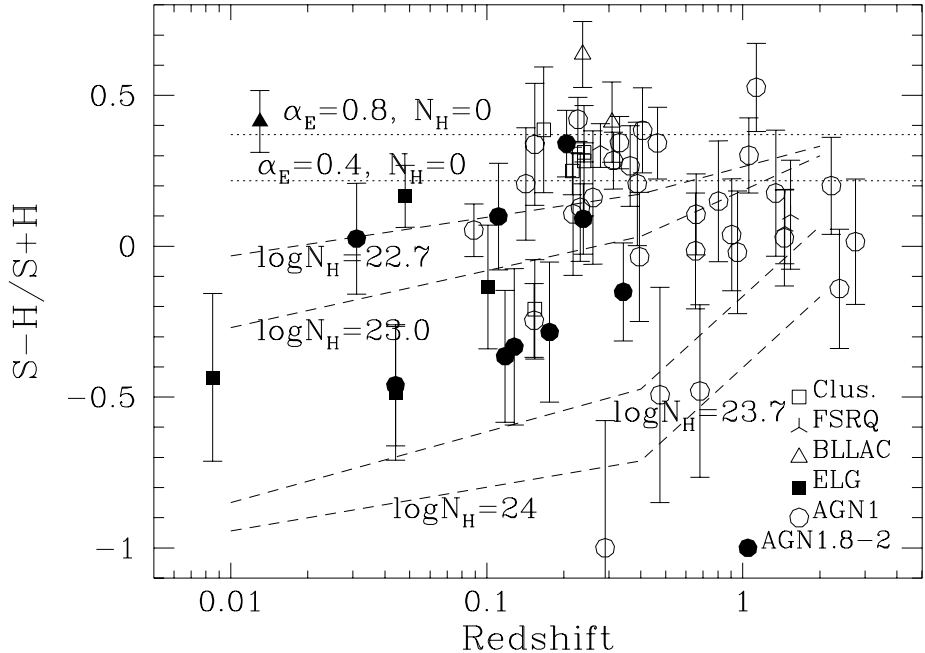}}
\vskip -0.3cm
\caption{The softness ratio (S--H/S+H) as a function of redshift
for the identified sources,
where S and H are the counts in the soft 1.3--4.5 and hard 4.5--10 
keV bands, respectively. The dotted lines show the expected hardness 
ratio for power law models, while 
dashed lines the expectations for absorbed power law models
with $\alpha$=0.8 and different values of column density at the source 
redshift.}
\end{figure}
%%%%%
In fact, while at redshifts lower than 0.3--0.4 X--ray obscured 
HELLAS sources are mainly associated to Type~2 AGNs and 
narrow emission-line galaxies, 
at higher redshifts the presence of X--ray absorption in a sizeable fraction 
of Type~1 AGNs represents a new, intriguing issue to be demanded to further 
{\it Chandra} and {\it XMM-Newton} observations, clearly 
suggesting that AGNs properties cannot be exhaustively interpreted without 
a multiwavelength approach. The decoupling between the optical and 
X--ray absorption 
properties of $z>$0.3--0.4 broad-line AGNs can be explained 
assuming a dust-to-gas ratio significantly lower than the Galactic value. 
From a physical perspective, this could imply the presence of a population 
of dust grains with different properties (e.g., large size) 
than previously known or thought (Maiolino et al. 2001). 
Strong X--ray absorption ($N_{\rm H}$ $>$ 10$^{23}$ cm$^{-2}$) 
in high-redshift broad-line AGNs 
has also been suggested to explain the X--ray properties of four objects 
detected in the course of the {\it ASCA} LSS (Akiyama et al. 2000), 
and recent results from {\it Chandra} 
observations seem to confirm this result at lower X--ray fluxes 
(Fiore et al. 2000, Akiyama et al. 2001). 
%Although X--ray absorption was previously detected in a large number 
%of radio-loud AGNs (e.g., Elvis et al. 1994; Cappi et al. 1997) 
%and in broad-absorption line quasars (e.g., Gallagher et al. 1999; 
%Brandt et al. 2000; Green et al. 2001), 
The discovery of X--ray absorption 
in optically ``normal'' broad-line AGNs has important consequences for the 
AGN synthesis models for the XRB.
According to these models, the sources 
responsible for the XRB must be characterized by a 
spectral energy density spanning a wide range of luminosities and 
absorption column densities, in order to reproduce both the XRB spectrum 
and the source counts in different energy ranges. 
In particular, the energetically dominant contribution comes from sources 
around the knee of the X--ray luminosity function 
($L_{\rm X}$ $\sim$ a few 10$^{44}$ erg s$^{-1}$ at z=1) and with 
absorbing column densities of the order of 10$^{23}$ cm$^{-2}$ 
(Comastri 2000). 
If one relies on the AGN unified model, 
these objects, the so-called ``QSO2'', are expected 
to be the luminous counterparts of the local Seyfert 2 galaxies 
with narrow optical emission lines. Despite extensive searches 
only a  handful of candidates have been found, the most remarkable
example being the $z$=3.7 quasar discovered in the 
{\it Chandra} deep field south (Norman et al. 2001).  
The results obtained by HELLAS suggest that if the 
statement: obscured Type~2 $\equiv$ narrow optical lines 
is not always true, the moderate/high-redshift absorbed Type~1 objects 
could have the same role of the so far elusive class of QSO2
in contributing to the XRB (Comastri et al. 2001). 

Absorption column densities in excess to the Galactic one, often larger than 
10$^{22}$--10$^{23}$ cm$^{-2}$, are also derived combining archival 
{\it ROSAT} data with the {\it BeppoSAX} ones (Vignali et al. 2001). 
Interesting enough, this broad-band X--ray analysis has revealed that 
a fraction of sources which are though to be absorbed in the hard X--rays 
are characterized by hard colors also in soft X--rays. 
This means that, although with lower efficiency, it is possible to pick up 
absorbed objects in the soft X--rays. It must be noted that some 
evidences of absorbed objects have also been obtained from the 
{\it ROSAT} Deep and Ultradeep Surveys in the Lockman Hole 
(Hasinger et al. 1998; Lehmann et al. 2000, 2001a). 
{\it ROSAT} observations of the HELLAS sources suggest 
that their average spectral properties are not well accounted for 
by a simple absorbed power-law model. 
More complex spectra are required in a high number of cases, 
and this is particularly true for very absorbed objects. 
Additional soft components possibly due to scattered nuclear 
flux and/or starburst components 
are required, in most cases involving about 10--50 \% of the 
primary radiation (Vignali et al. 2001). 
Similar results have been obtained by {\it ASCA} (Della Ceca et al. 1999). 

Remarkable results have been obtained also by the photometric follow-up 
observations of a subsample of HELLAS sources. Interesting enough, 
the optical and the near-infrared properties of a fraction of intermediate 
(1.8--1.9) and Type~2 objects (Maiolino et al. 2000), 
as well as red quasars (Vignali et al. 2000), are dominated by 
the stellar component of the galaxies hosting the (obscured) 
X--ray active nucleus. 
This result has straightforward consequences, since it implies that 
a fraction of sources responsible for the hard XRB may be hosted 
by normal, passively-evolving galaxies, possibly being missed 
by previous optical surveys based on color-selection criteria.

\section{From BeppoSAX to XMM--Newton: The HELLAS2XMM survey}

The {\tt HELLAS2XMM} survey aims to cover a portion of the 
redshift--luminosity plane which cannot be probed 
by deep pencil-beam surveys.
The main purpose of this complementary approach is to study the 
X--ray source 
populations at fluxes where a large fraction of the hard X-ray cosmic
background (HXRB) is resolved ($\approx50\%$ at $F_{2-10}>10^{-14}$
\cgs, see e.g. Comastri 2000), but where a) the area
covered is as large as possible, to be able to find sizeable samples
of ``rare'' objects; b) the X-ray flux is high enough to provide at
least rough X-ray spectral information; and c) the magnitude of the
optical counterparts is bright enough to allow, at least in the
majority of the cases, relatively high-quality optical spectroscopy,
useful to investigate the physics of the sources.
Our goal is to evaluate for the first time the luminosity function of
hard X-ray selected sources in wide luminosity and redshift ranges.
By integrating this luminosity function we will compute the hard
X-ray luminosity density per unit volume due to accretion as a
function of the redshift. A comparison with the history
of the UV luminosity density (proportional to the history of the
star-formation)  may give us a clue on the
correlations between formation and evolution of AGN and supermassive
black holes and formation and evolution of galaxies.

At present, the {\tt HELLAS2XMM} sample, performed using 15 
XMM--{\it Newton} public observations, 
consist of 1022, 495, and 100 sources detected down to minimum fluxes of 
about 6 $\times$ 10$^{-16}$, 3 $\times$ 10$^{-15}$, 6 $\times$ 10$^{-15}$
\cgs in the 0.5--2, 2--10 and 4.5--10 keV bands, respectively, over an area 
of about 3 deg$^2$ (Baldi et al. 2001). The source counts in these bands      
are in good agreement with previous determination by other satellites
and XMM--{\it Newton} itself. In the hard 2--10 and 5--10 keV bands
our survey samples a flux range neither accessible by shallower
{\it ASCA} (Ueda et al. 1999, Della Ceca et al. 1999)
and {\it BeppoSAX} survey (Fiore et al. 2001a) which were limited to
relatively bright fluxes, 
nor by deep {\it Chandra} (Brandt et al. 2001, Rosati et al. 2001)
and XMM--{\it Newton} (Hasinger et al. 2001) 
surveys which are limited by the small area. 

Four of the {\tt HELLAS2XMM} fields were selected for follow--up observations 
in the optical band using the ESO 3.6m and the TNG 3.5m telescopes.
The selected subsample contains 115 sources with 2--10 keV fluxes
between $\sim 8 \times$ 10$^{-15}$ and  10$^{-13}$ \cgs.
Medium--deep R band images are available for all the sources in the 
sample. Optical counterparts brighter than R$\simeq$24 
within 5 arcsec from the X--ray position 
(actually within 3 arcsec in most of the cases; see Fiore et al. 2001b) 
are present for about 80\% of the sample.
At the time of writing optical spectra have been obtained 
for 46 out of the 115 sources in our sample. 
The number of faint hard X--ray selected sources with optical identification 
is comparable to that obtained in deep {\it Chandra} observations
(Barger et al. 2001). 
The {\tt HELLAS2XMM} 
sources populate a region of the luminosity--redshift diagram
which is barely covered by deep surveys (Fig.~2).
A uniform sampling over a large region of the $L_X-z$ parameter space 
is a key issue to compute the luminosity density and evolution 
of the X--ray sources. 
%%%%%%
\vskip -1.3cm
\begin{figure}[h]
%\plotone{fig2.ps}
\centerline{\includegraphics[angle=0,width=0.7\textwidth]{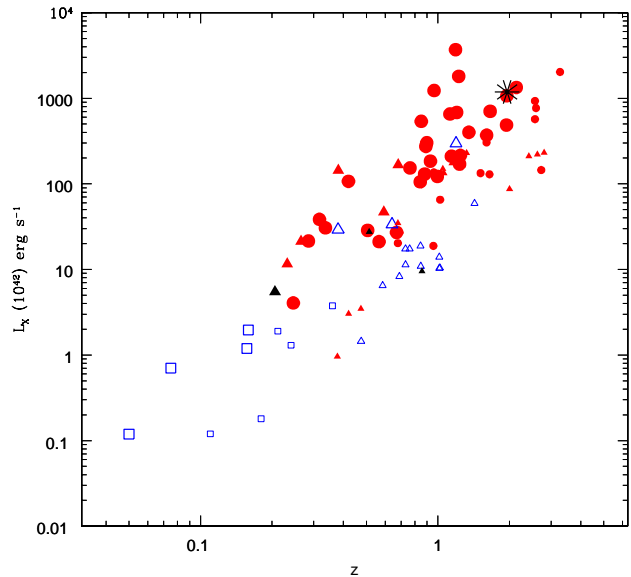}}
\vskip -0.3cm
\caption{The luminosity-redshift diagram for the HELLAS2XMM sources 
(big symbols) and the {\it Chandra} SSA13, HDFN and A370 
deep surveys (smaller symbols, data from Barger et al. 2001).
Different symbols identify different source classes:
filled circles = broad-line quasars and Sy1; 
filled triangles = narrow-line AGN; 
open squares = optically `normal' galaxies; 
open triangles = emission-line galaxies; big star = the candidate type 2 QSO at $z$=1.955.}
\end{figure}
%%%%%%

\subsection{Optical identification breakdown}

The most interesting results emerging from the optical identification 
program are the discovery of one X--ray luminous ($L_X \simeq$ 10$^{45}$
erg s$^{-1}$) quasar at $z$ = 1.955 with relatively 
narrow optical emission lines 
(FWHM $\ls$  1000 km s$^{-1}$; FWZI $\ls$ 2000 km s$^{-1}$ in the rest frame)
and of some X--ray luminous but apparently 
normal galaxies at low redshift. 
Although type 2 quasars were predicted by AGN synthesis model 
for the XRB, their space density and optical appearance 
is still matter of debate (see the previous paragraph).
The present finding suggests that hard X--ray selection 
coupled with large area provides 
an efficient  method to uncover new objects of this class.
More surprising is the presence of X--ray bright sources in the nuclei 
of otherwise passive ``normal'' galaxies.  
Their 2--10 keV luminosities, in the range 10$^{42-43}$ erg s$^{-1}$, 
are more than one order of magnitude higher than that predicted on 
the basis of the $L_X$--L$_{opt}$ relation of normal galaxies 
(Fabbiano et al. 1992) and more typical of Seyfert galaxies. 
Their count ratios indicate a hard X--ray spectrum.
A detailed broad-band study has been recently performed on one 
of these objects (Comastri et al. 2002) suggesting that 
an heavily obscured, Compton-thick \hbox{($N_H > 10^{24}$ cm$^{-2}$)} AGN 
may be responsible for the observed properties (Fig.~3).  
%%%%%%%
\vskip -1.3cm
\begin{figure}[h] 
%\plotone{sed_6240.ps}
\centerline{\includegraphics[angle=0,width=0.7\textwidth]{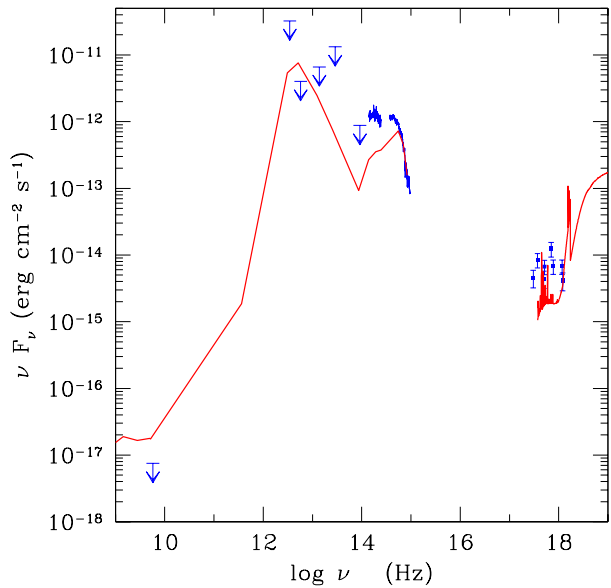}}
\vskip -0.3cm
\caption{The spectral energy distribution (SED) 
of the X--ray bright optically
quiet galaxy Fiore P3 (see Comastri et al. 2002 for details) 
is compared with that of the highly obscured Seyfert 2 galaxy NGC 6240
(solid line). The latter SED was normalized to the observed optical spectrum.}
\end{figure}
%%%%%%

\subsection{X--ray to optical properties}

The results of photometric observations indicate that $\sim$ 23 \% of the 
X--ray sources have counterparts with $22 \ls R \ls 24$ and 
$\sim$ 17 \% with R $>$ 24. The identification of the optically faintest 
counterparts will not probably be feasible even with 8m class telescopes
requiring deep multifilter observation to get reliable photometric redshifts.
An alternative solution would be to search for redshifted iron 
K$\alpha$ lines. Although such an approach may be feasible for the 
brightest X--ray sources, for the large majority of the objects 
in our sample the X--ray spectrum cannot be measured 
due to the low counting statistics. 
Useful constraints on the nature of faint X--ray source population 
might be obtained from the analysis of the already available  
optical and X--ray fluxes and from an estimate of their average X--ray
spectral properties inferred from the hardness ratio analysis. 
The R band magnitudes plotted versus the 2--10 keV flux are reported
in Fig.~4. 
%%%%%%
%\vskip -1.8cm
\begin{figure}[t]%[h]
%\plotone{fig4.ps}
\centerline{\includegraphics[angle=0,width=0.7\textwidth]{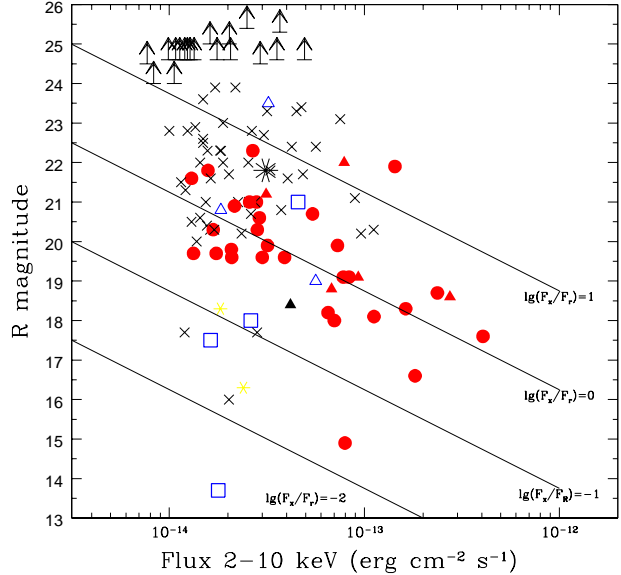}}
\vskip -0.3cm
\caption{The 2--10 keV X--ray flux plotted against the
R magnitude. The symbols are the same of Fig.~2, while the crosses represent
unidentified sources. Two Galactic stars are also plotted as asterisks. 
The locii of constant $F_X/F_{opt}$ are reported with the values as labeled.}
\end{figure}
%%%%%%
The so far identified AGN show a relatively well-defined correlation 
with the optical magnitude around $F_X/F_{opt} \simeq$ 1.
This correlation is similar to that found by {\it ROSAT} for soft X--ray selected 
quasars (Hasinger et al. 1998) and confirmed by {\it Chandra} and
XMM--{\it Newton} observations (Lehmann et al. 2001b) also for
hard X--ray selected sources.
On the other hand, the X--ray to optical flux ratio of unidentified sources  
is characterized by a larger scatter and skewed towards
higher $F_X/F_{opt}$ values. At faint fluxes 
there are several objects with $F_X/F_{opt}>$ 10 
suggesting the presence of highly obscured AGN.
In order to quantify this possibility we have computed
the average hardness ratio as a function of the X--ray to optical flux ratio
and the X--ray flux. The entire sample has been divided into bright 
and faint sources according to the median flux of the 
survey in the hard X--ray band: 2.5 $\times$ 10$^{-14}$ \cgs.
%%%%%%
\begin{figure}[h]%[t]
%\plotone{fig5.ps}
\centerline{\includegraphics[angle=0,width=0.7\textwidth]{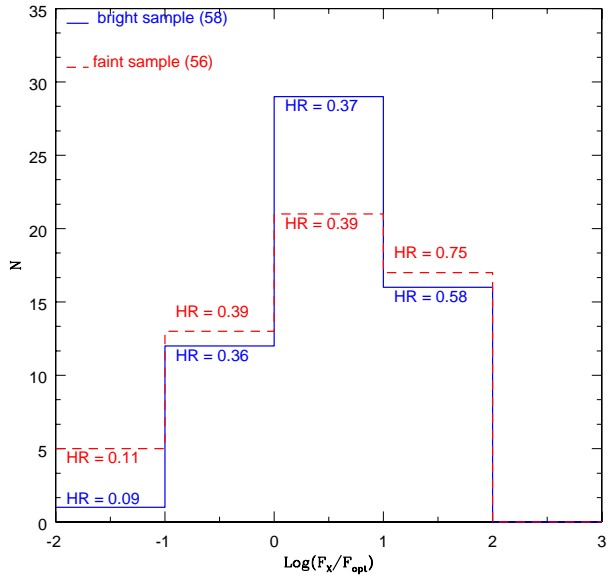}}
\caption{The distribution of X--ray to optical flux ratios for the 
bright and faint samples. The labeled values correspond to the average 
hardness ratio in the bin defined as H--S/H+S where H and S are the 
2--10 keV and 0.5--2 keV fluxes respectively. 
Sources with log$F_X/F_{opt}>$ 1 are grouped in a single bin.}
\end{figure}
%%%%%%

The results reported in Fig.~5 indicate a hardening 
of the X--ray spectrum for those 
sources with the highest values of $F_X/F_{opt}$, this effect being 
more pronounced in the faint sample.

\section{Conclusions}

The relatively high number of obscured AGN 
discovered by {\it BeppoSAX} and XMM--{\it Newton} 
makes high-energy large area surveys extremely well suited to 
study the physics and the evolution of the sources responsible for the 
hard X--ray background.
The hard X--ray sky is populated by AGN with extremely varied
broad-band properties.
The most important result concerns the optical appearance of X--ray
obscured AGN as a function of redshift.
In the local Universe a fraction of absorbed objects are associated with 
apparently normal, early-type galaxies. The lack of any AGN 
feature in their optical--infrared spectra suggests the presence 
of buried, probably Compton-thick nuclei.
At higher redshift the presence of hard X--ray sources with 
broad optical lines indicates that the absorbing gas 
is dust--free. Finally a sizeable fraction of hard X--ray selected sources 
lacks an optical counterpart at the limit of 4m class telescopes.
Multiwavelength observations of hard X--ray selected 
sources allow us to uncover AGN activity in a number of objects
which would have not been classified as such on the basis 
of observations at other wavelengths.
Larger samples of hard X--ray selected sources will provide new insights 
into the physics and the cosmic history of accretion.

\acknowledgments
This research has been partially supported by ASI contracts
ARS--99--75 and I/R/107/00, and by the MURST grant Cofin-00--02--36. 
CV also aknowledges the financial support of NASA LTSA grant NAG5-8107.

\end{document}